 \documentclass[prc,floatfix,onecolumn,superscriptaddress]{revtex4}
\newcommand{\GPDtE}{\langle \tilde{E} \rangle}
\newcommand{\GPDtH}{\langle \tilde{H} \rangle}

\newcommand{\GPDHT}{\langle H_T \rangle}
\newcommand{\GPDETbar}{\langle \bar{E}_T \rangle}

\newcommand{\HT}{\langle H_T \rangle}
\newcommand{\ETbar}{\langle \bar{E}_T \rangle}
\usepackage{amsmath}
\usepackage{graphicx}
\usepackage{url}
\usepackage{color}
\usepackage{subfigure}
\usepackage{longtable}
\usepackage{dcolumn}

\begin{document}

\markboth{V.Kubarovsky}
{Deeply Virtual Meson Production and Transversity GPDs}

%
%

\title{Deeply Virtual Pseudoscalar Meson Production at Jefferson Lab and Transversity GPDs}

\author{Valery Kubarovsky}

\address{Thomas Jefferson National Accelerator Facility\\
Newport News, VA 23606,
USA\\
vpk@jlab.org\\
and the CLAS Collaboration}

\begin{abstract}

The cross section of the exclusive $\pi^0$ and $\eta$ electroproduction reaction $ep\to e^\prime p^\prime \pi^0/\eta$ was measured at Jefferson Lab with a 5.75-GeV electron beam and the CLAS detector. Differential cross sections $d^4\sigma/dtdQ^2dx_Bd\phi$  and structure functions $\sigma_T+\epsilon\sigma_L, \sigma_{TT}$ and $\sigma_{LT}$ as functions of $t$ were obtained over a wide range of $Q^2$ and $x_B$. The data are compared with the GPD based theoretical models. Analyses find  that a large dominance of transverse processes is necessary to explain the experimental results. 
Generalized form factors of the transversity GPDs $\HT^{\pi,\eta}$ and $\ETbar^{\pi,\eta}$ were directly extracted from the experimental observables
for the first time. It was found that GPD $\bar E_T$ dominates in  pseudoscalar meson production.
The combined $\pi^0$ and $\eta$ data opens the way for the flavor decomposition of the transversity GPDs. 
The first ever evaluation of this decomposition was demonstrated.

\end{abstract}
\maketitle

\section{Introduction}	

Understanding nucleon structure in terms of the fundamental degrees of freedom of Quantum Chromodynamics (QCD) is one of the main goals in the theory of strong interactions.   In recent years it became clear that  exclusive reactions may provide information 
about hadron structure
encoded in so-called Generalized Parton Distributions  \cite{Ji,Radyushkin} (GPDs).
For each quark flavor $q$
there are eight GPDs. Four correspond to  parton helicity-conserving (chiral-even) processes,  denoted 
by $H^q$,  $\tilde H^q$,  $E^q$ and  $\tilde E^q$, and 
four correspond to parton helicity-flip (chiral-odd) processes  \cite{ji,diehl},  $H^q_T$,  $\tilde H^q_T$,  $E^q_T$ and  $\tilde E^q_T$. 
The GPDs depend on three kinematic variables: $x$, $\xi$ and $t$. In  a symmetric frame,  $x$ is the   average longitudinal momentum fraction of the struck parton before and after the hard interaction and $\xi$ (skewness) is half of the longitudinal momentum fraction transferred to  the struck parton. The skewness can be expressed in terms of the   Bjorken variable $x_B$  as
$\xi\simeq x_B/(2-x_B)$. Here $x_B=Q^2/(2p\cdot q)$ and $t=(p-p^\prime)^2$, where $p$ and $p^\prime$ are the initial and final four-momenta of the nucleon.

When the theoretical calculations for longitudinal virtual photons were compared with the JLab  $\pi^0$ data\cite{clas1,clas2}  
they were  found  to  underestimate the measured cross sections by more than an order of magnitude in their accessible kinematic regions.
The failure to describe the experimental results with quark helicity-conserving operators  stimulated a consideration of the role of the  chiral-odd quark helicity-flip processes. Pseudoscalar meson electroproduction, and in particular $\pi^0$ production in the reaction $ep\to e^\prime p^\prime \pi^0$, was 
identified~\cite{Ahmad:2008hp,G-K-09,G-K-11} 
as especially sensitive to the quark helicity-flip subprocesses. 
During the past few years, two parallel theoretical approaches - \cite{Ahmad:2008hp,GL}~(GL) and \cite{G-K-09,G-K-11}~(GK) have been developed utilizing the  chiral-odd GPDs in the calculation of  pseudoscalar meson electroproduction. The GL and GK approaches, though employing different models of  GPDs, lead to {\it transverse} photon amplitudes that are much larger than the longitudinal amplitudes.

\section{Definition of Structure Functions}

The unpolarized reduced meson cross section is described 
 by 4 structure functions $\sigma_T$, $\sigma_L$, $\sigma_{TT}$ and $\sigma_{LT}$ 
\begin{align}
2\pi\frac{d^2\sigma(\gamma^*p\to p\pi^0)}{dtd\phi_\pi} &= 
\frac{d\sigma_T}{dt} + 
\epsilon  \frac{d\sigma_L} {dt}+ 
\epsilon  \frac{d\sigma_{TT}} {dt}  \cos 2\phi+
\sqrt{2\epsilon (1+\epsilon)}   \frac{d\sigma_{LT}} {dt}  \cos \phi \notag.
\end{align}

\noindent 
References~\cite{G-K-11,GL} obtain the following relations for unpolarized structure functions:

\begin{align}
\label{SL}
\frac{d\sigma_{L} }{dt}&= \frac{4\pi\alpha}{k^\prime}\frac{1}{Q^4}\left\{ \left( 1-\xi^2 \right) \left|\GPDtH\right|^2 -2\xi^2\text{Re}\left[ \GPDtH^* \GPDtE \right] - \frac{t^\prime}{4m^2} \xi^2 \left| \GPDtE \right|^2 \right\},
\end{align}

\begin{align}
\label{ST}
\frac{d\sigma_{T}}{dt} &= \frac{4\pi\alpha}{2k^\prime Q^4} \left[ \left(1-\xi^2\right) \left|\GPDHT\right|^2 - \frac{t'}{8m^2} \left|\GPDETbar\right|^2\right],
\end{align}

\begin{align}
\label{SLT}
\frac{d\sigma_{LT}}{dt} = \frac{4\pi\alpha}{\sqrt{2}k^\prime Q^4} \xi \sqrt{1-\xi^2} \frac{\sqrt{-t'}}{2m} \text{ Re} \left[ \langle H_T\rangle^* \langle\tilde{E}\rangle \right],
\end{align}

\begin{align}
\label{STT}
\frac{d\sigma_{TT}}{dt} = \frac{4\pi\alpha}{k^\prime Q^4}\frac{t'}{16m^2}\left|\GPDETbar\right|^2,
\end{align}

\noindent 
Here $m$ is the mass of the proton, $t^\prime =t-t_{min}$, where $|t_{min}|$ is the minimum value of $|t|$ corresponding to $\theta_\pi =0$, $k^\prime(Q^2,x_B)$ is a phase space factor and  $\bar E_T = 2\widetilde H_T + E_T$.
The brackets $\langle  H_T \rangle$ and $\langle \bar E_T \rangle$ denote
the convolution of the elementary process 
$\gamma^*q\to q\pi^0$
with the GPDs $H_T$ and $\bar E_T$. We call them generalized form factors.

\begin{figure}[h]
\centerline{
\includegraphics[width=10cm]{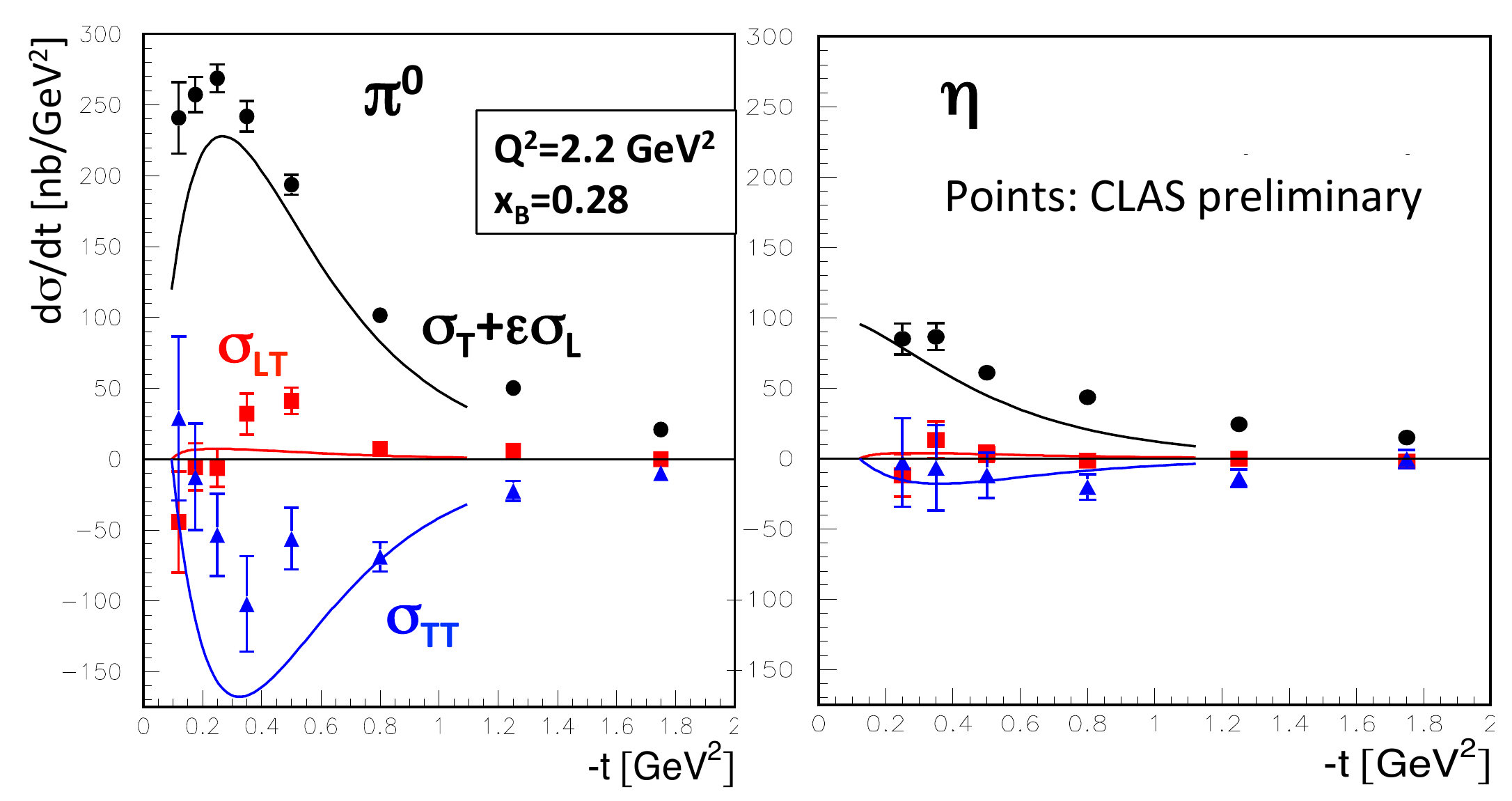}
}
\caption
{
(Color online) Structure functions $\sigma_T+\epsilon\sigma_L$ (black), $\sigma_{TT}$ (blue) and $\sigma_{LT}$ (red) as a function
of -$t$ for $\pi^0$ (left) and $\eta$(right) exclusive electroproduction for kinematic point ($Q^2=2.2$~GeV$^2$,$x_B$=0.28).
Data points: CLAS, preliminary. Curves: theoretical predictions produced with the GK handbag model.
}
\label{fig:structure}
\end{figure}

\section{Experimental data}
Cross section of the reaction $ep\rightarrow ep\pi^0$ measured by 
the CLAS collaboration at Jlab at
 1800 kinematic points in bins of $Q^2$, $x_B$, $t$ and $\phi$ were published in 
 Refs.
 \cite{clas1} and \cite{clas2}.
Structure functions $\sigma_U=\sigma_T+\epsilon\sigma_L$, $\sigma_{LT}$ and $\sigma_{TT}$ have been obtained. These functions were compared 
with the predictions of the GPD models \cite{G-K-11,GL}. CLAS confirmed that the measured unseparated cross sections are much larger than expected from 
leading-twist handbag calculations which are dominated by longitudinal photons. The same conclusion can be made in an almost model-independent way by noting that the structure functions $\sigma_U$ and $\left| \sigma_{TT}\right|$ are comparable to each other while $\left| \sigma_{LT}\right|$ is quite small (see Fig.\ref{fig:structure}).
Cross section and structure functions for $ep\to e \eta p$ were also obtained in parallel.
The comparison of the $\pi^0$ and preliminary $\eta$ structure functions is shown in Fig.~\ref{fig:structure}. 
$\sigma_U$ drops by a factor of 2.5 for $\eta$ in comparison with $\pi^0$ and
$\sigma_{TT}$ drops by a factor of 10.
The GK GPD model \cite{G-K-11}  (curves)  follows the experimental data.
The inclusion of $\eta$ data into consideration strengthens 
the statement about the transversity GPD dominance in the pseudoscalar electroproduction process.


\section{Generalized Form Factors}
\begin{figure}[h]
\vspace*{-10 mm}
\centerline{
\includegraphics[width=16cm]{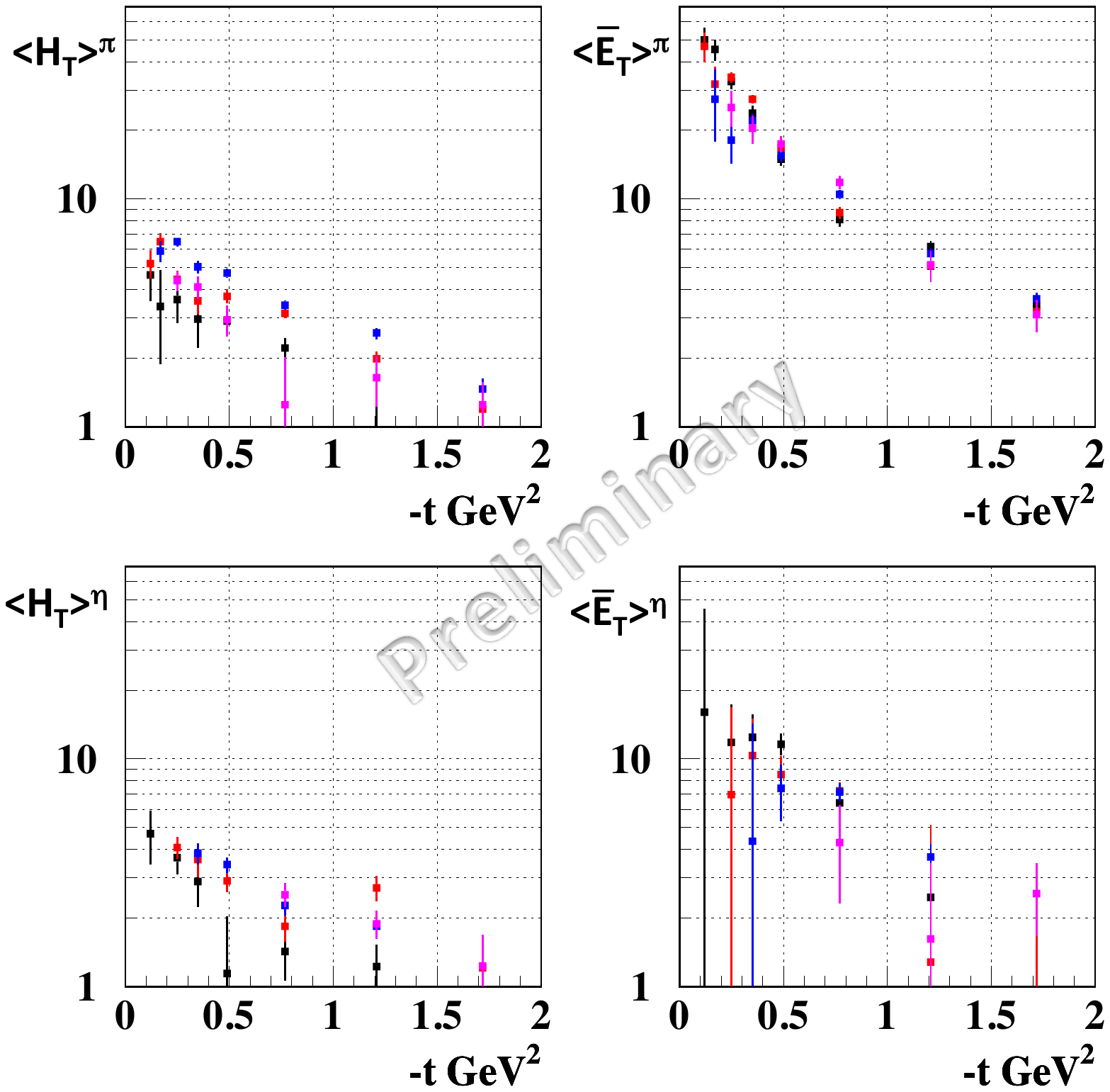}
}
\vspace*{-15mm}
\caption{(Color online) Data points: CLAS, preliminary.
Top left: $\left|\GPDHT^\pi\right|$,
top right: $\left|\GPDETbar^\pi\right|$,
bottom left: $\left|\GPDHT^\eta\right|$,
bottom right: $\left|\GPDETbar^\eta\right|$ as a function of -$t$ for different values of 
($Q^2$~$[GeV^2]$,$x_B$)=(1.2,~0.15) black, (1.8,~0.22) red, (2.2,~0.27) blue, (2.7,~0.34) magenta.
}
\label{fig:ht_et_pi0_eta}
\end{figure}

The squared magnitudes of the 
generalized form factors
$\left|\GPDHT\right|^2$ and 
$\left|\GPDETbar\right|^2$  
may be directly extracted from the experimental data (see Eqs. \ref{ST} and \ref{STT} ) in the framework of GPD models.
\begin{equation}\label{ET_HT}
\begin{aligned}
\left|\GPDETbar^{\pi,\eta}\right |^2&=\frac{k^\prime Q^4}{4\pi\alpha}     \frac{16m^2}{t'} \frac{d\sigma^{\pi,\eta}_{TT}}{dt} \\
\left| \GPDHT^{\pi,\eta}\right|^2&=\frac{2k^\prime Q^4}{4\pi\alpha}       \frac{1}{1-\xi^2} \left [ \frac{d\sigma^{\pi,\eta}_{T}}{dt} + \frac{d\sigma^{\pi,\eta}_{TT}}{dt} \right ].
\end{aligned}
\end{equation}
\noindent
Fig.~\ref{fig:ht_et_pi0_eta} presents the modules of the generalized form factors
$\left|\GPDHT^\pi\right|$, $\left|\GPDETbar^\pi\right|$, $\left|\GPDHT^\eta\right|$ and $\left|\GPDETbar^\eta\right|$
 for 4 different kinematics. 
  Note the dominance of the $\left|\GPDETbar\right|$ over $\left|\GPDHT\right|$ for both $\pi^0$ and $\eta$ for all kinematics.

\begin{figure}[h]
\centerline{
\includegraphics[width=10cm]{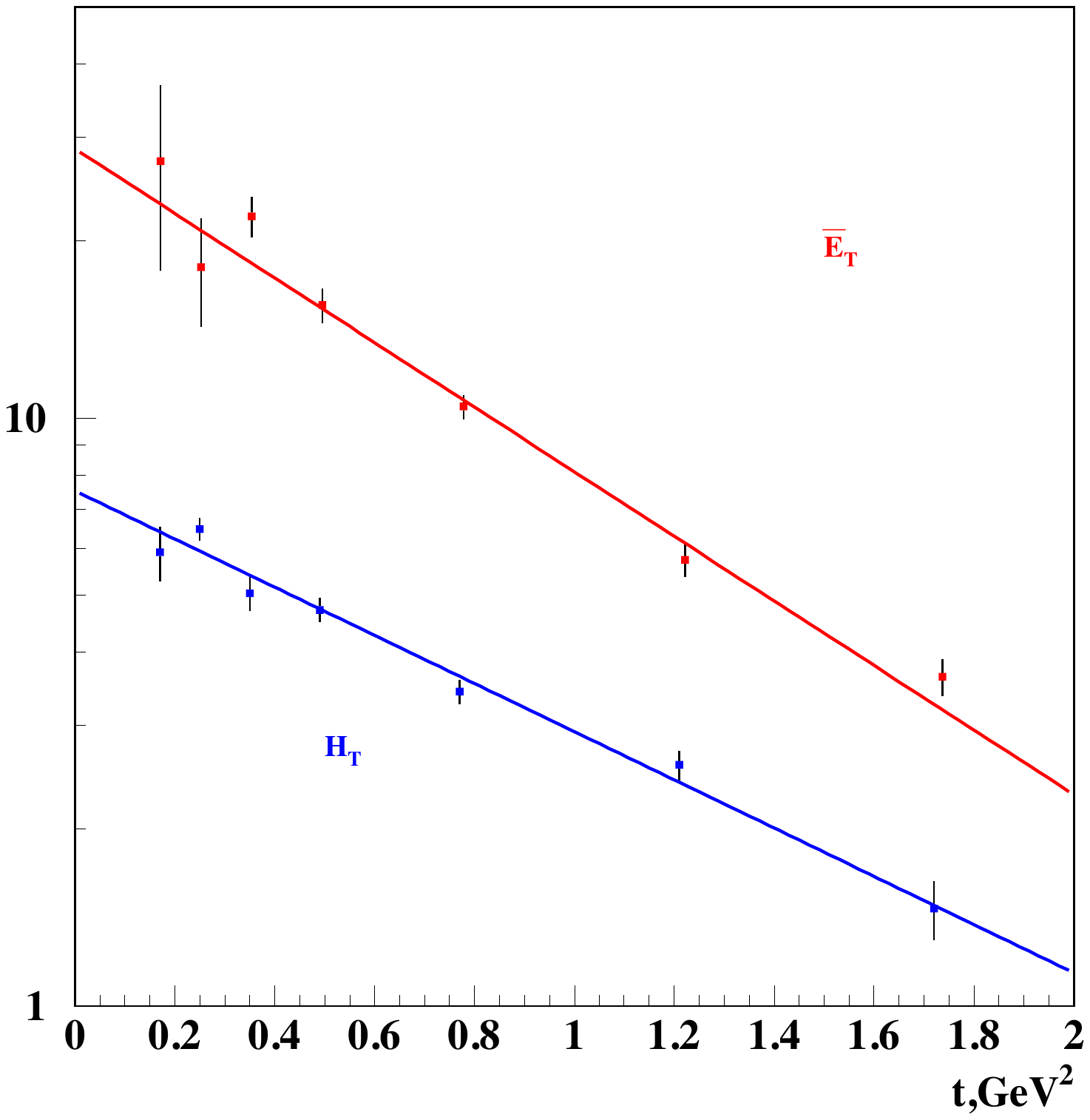}
}
\vspace*{-30 mm}
\caption
{(Color online) Preliminary.
$\left|\GPDHT^\pi\right|$ (blue) and
$\left|\GPDETbar^\pi\right|$ (red),
as a function of -$t$, ($Q^2=2.2$ GeV$^2$,  $x_B$=0.27).
}
\label{fig:ht_et_pi0_3}
\end{figure}

Generalized formfactors $\left|\GPDHT^\pi\right|$ and $\left|\GPDETbar^\pi\right|$ 
may be fitted by simple exponential function with good $\chi^2$.
The result is shown in Fig.~\ref{fig:ht_et_pi0_3} for one of the kinematical point.  Note that $\left|\GPDETbar^\pi\right|$  form factor has steeper t-slope than
$\left|\GPDHT^\pi\right|$. This is the first attempt to evaluate generalized formfactors from the experimental data. 

\begin{figure}[pb]
\centerline{
\includegraphics[width=6.7cm]{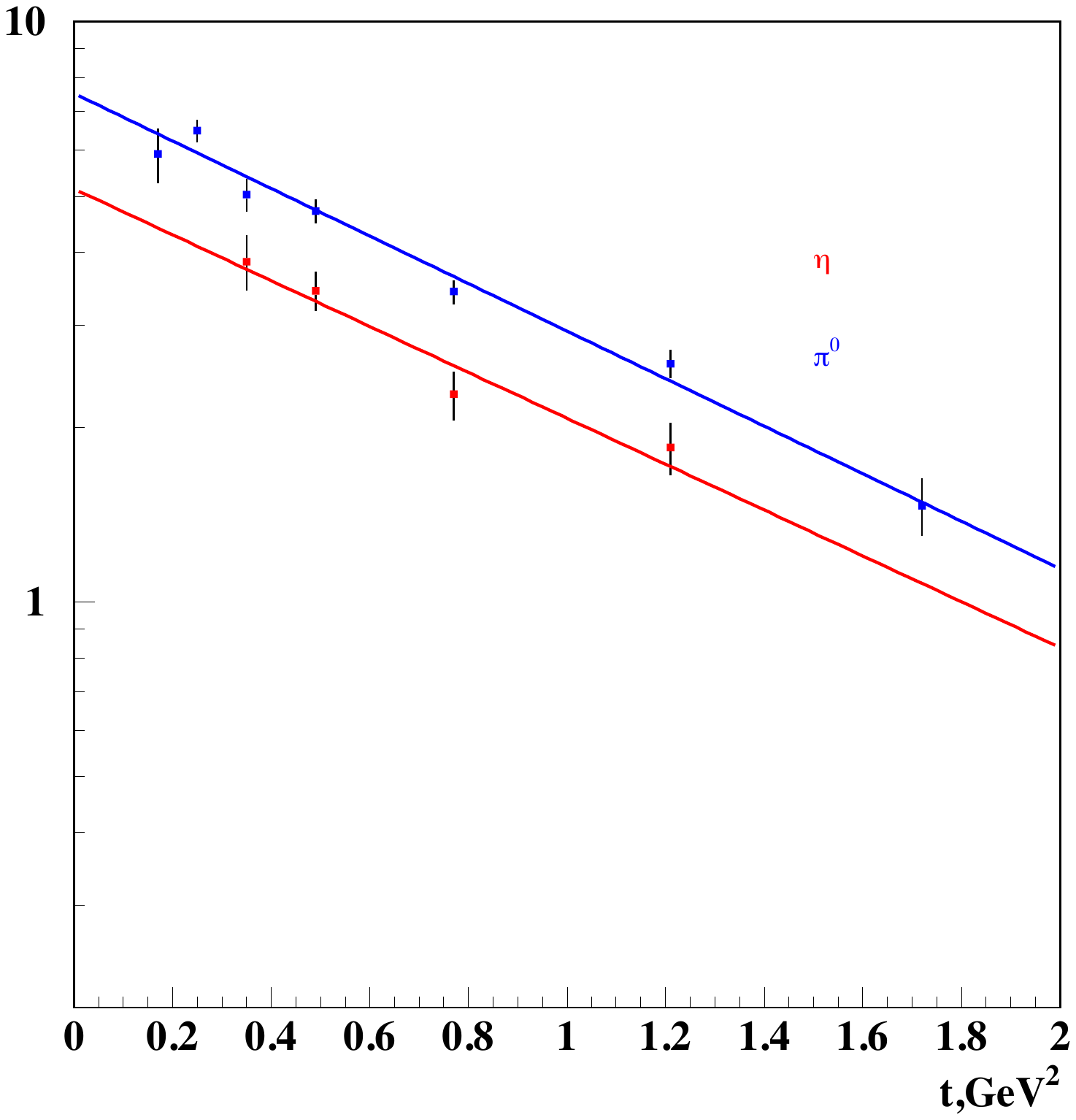}
\includegraphics[width=6.7cm]{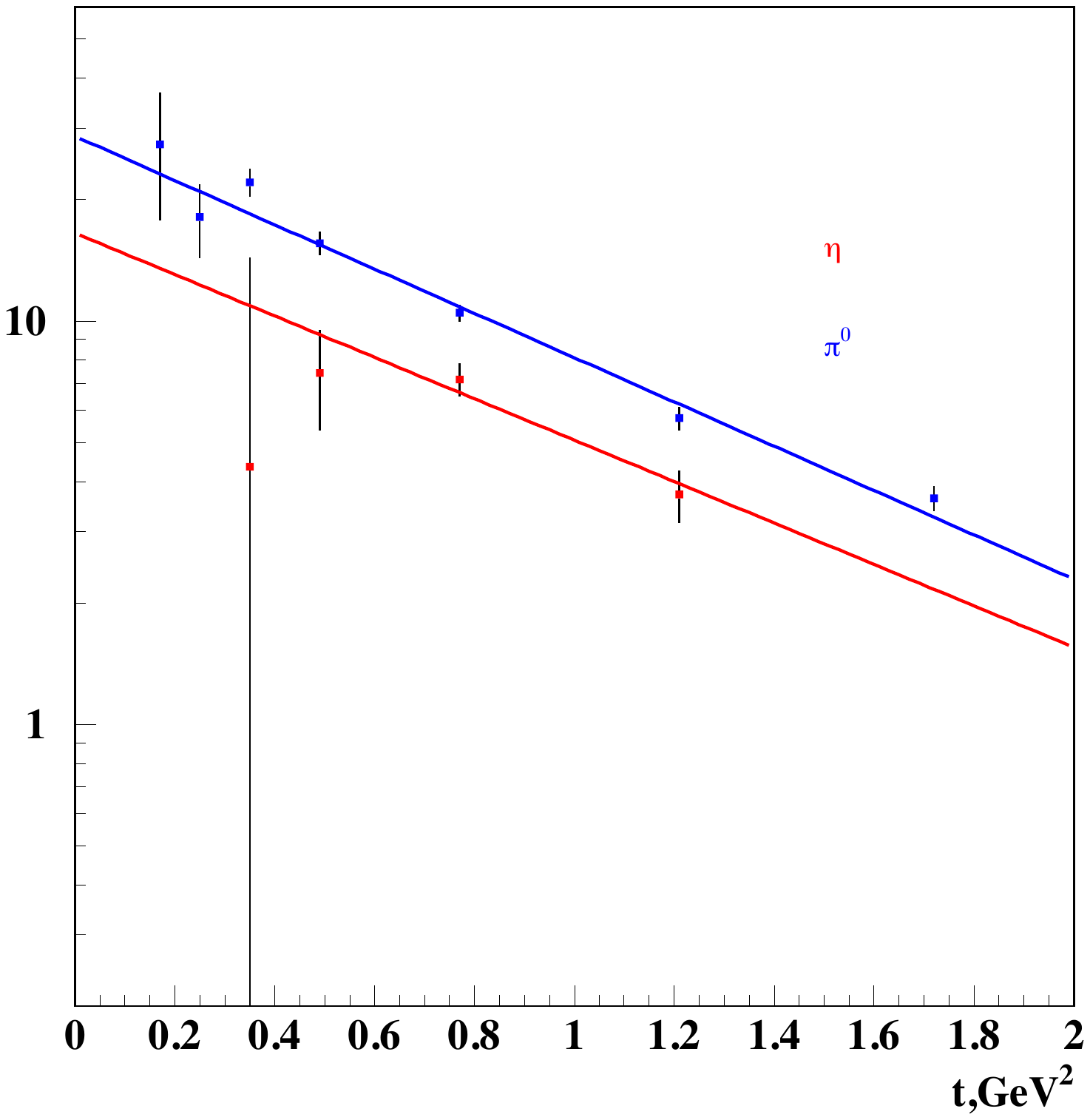}
}
\vspace*{-10 mm}
\caption
{(Color online) Preliminary.
Left:   $\left|\GPDHT^\pi\right|$(blue) and $\left|\GPDHT^\eta\right|$(red).
Right: $\left|\GPDETbar^\pi\right|$(blue) and $\left|\GPDETbar^\eta\right|$(red);
($Q^2$=2.2, $x_B$=0.27).
}
\label{fig:ht_pi0_eta}
\end{figure}

Fig.~\ref{fig:ht_pi0_eta} presents the comparison of the generalized form factors  $\left|\GPDHT\right|$ and  $\left|\GPDETbar\right|$ for $\pi^0$ (blue) and $\eta$ (red)
for oner kinematical point ($Q^2$=2.2, $x_B$=0.27).
These data will be used for the flavor decomposition in the next section.

\section{Flavor Decomposition}

In  electroproduction the GPDs $F_i$ appears in the following combinations
\begin{equation}
\begin{aligned}
F^\pi_i&=\frac{1}{\sqrt{2}}[e_uF^u_i-e_dF^d_i]\\
F^\eta_i&=\frac{1}{\sqrt{6}}[e_uF^u_i+e_dF^d_i-2e_sF^s_i]\\
\end{aligned}
\end{equation}
\noindent
The $q$ and $\bar q$ GPDs contribute in the quark combinations $F^q_i-F^{\bar q}_i$. 
Hence there is no contribution from the strange quarks if we assume that
$F^s_i\simeq F^{\bar s}_i$. 
For flavor decomposition we have to take into account the decay constants $f_\pi$ and $f_\eta$, the chiral  condensate
constants $\mu_{\pi^0}=$2.57~GeV, $\mu_1=$0.958~GeV and $\mu_8$=2.32~GeV, and the contribution from singlet and octet $\eta$ states.\cite{G-K-11}

\begin{equation}
\begin{aligned}
F^\eta_i=F^{\pi}_i\left(\cos\theta_8-\sqrt{2} \frac{\mu_1}{\mu_8}\frac{f_1}{f_8}\sin\theta_1\right)\frac{f_8}{f_{\pi^0}}\frac{\mu_8}{\mu_{\pi^0}}=\frac{ F^8_i}{k_\eta},
\end{aligned}
\end{equation}
\noindent
where the mixing angles are:
$\theta_8=-21.2^o$ and $\theta_1=-9.2^o$.
The octet and singlet wave functions are very similar and the decay constants are close as well
$f_8=1.26f_\pi$ and 
$f_1=1.17f_\pi$.
The overall factor for the $\eta$ meson is 
$k_\eta=0.863$.
Using $e_u=\frac{2}{3}$ and $e_d=-\frac{1}{3}$ we will end up with
equations
\begin{equation}\label{decom}
\begin{aligned}
F^\pi_i&=\frac{1}{3\sqrt{2}}[2F^u_i+F^d_i]\\
{k_\eta}{F^\eta_i}&=\frac{1}{3\sqrt{6}}[2F^u_i-F^d_i].
\end{aligned}
\end{equation}
\noindent
\noindent
Experimentally we have  access only to the $\left|\left<F_i^\pi\right>\right|^2$ and $\left|\left<F_i^\eta\right>\right|^2$
(see Eq.~\ref{ET_HT}).
The final equation for the $\left<H_T\right>$ convolution reads

\begin{equation}
\left\{
\begin{aligned}
\frac{1}{18}       \left | 2\HT^u + \HT^d  \right  |^2 &=  \left |\GPDHT^\pi \right |^2\\
 \frac{1}{54}       \left  | 2\HT^u - \HT^d  \right  |^2 &= k_\eta^2\left |\GPDHT^\eta \right |^2
\end{aligned}
\right.
\end{equation}
\noindent
and simular equations for $\ETbar$.
\begin{figure}[h]
\vspace*{-10 mm}
\centerline{
\includegraphics[width=12cm]{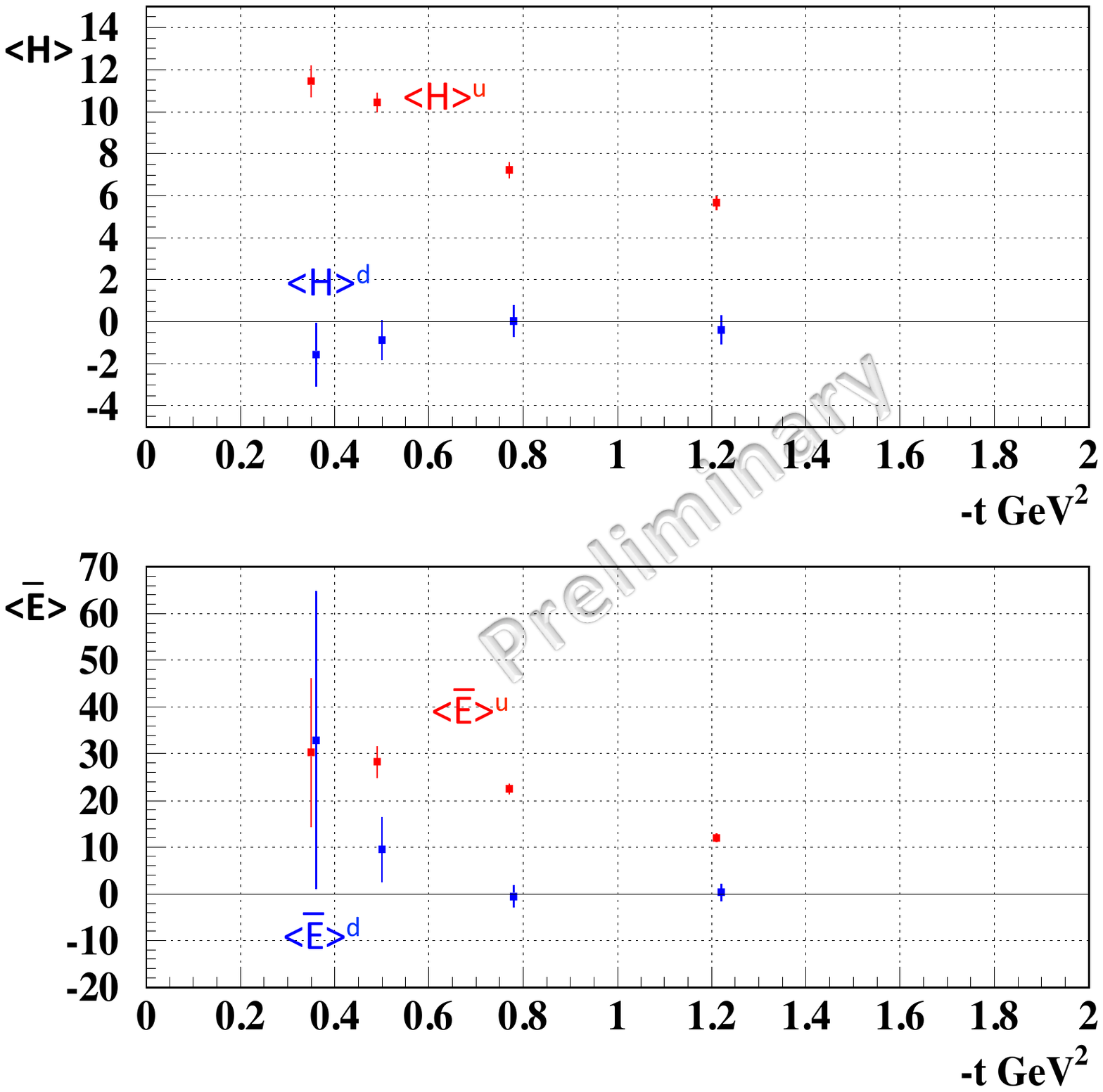}
}
\vspace*{-10 mm}
\caption
{
(Color online) Preliminary. 
Top: Extracted $\GPDHT^u$ (red) and $\GPDHT^d$ (blue);
Bottom: $\GPDETbar^u$ (red) and $\GPDETbar^d$ (blue),
as a function of -$t$ for $Q^2=2.2~GeV^2$  and $x_B$=0.27.
}
\label{fig:ud_3}
\end{figure}
\noindent
The solution of these equations will lead to the flavor decomposition of the transversity GPDs 
$\HT^u$ and  $\HT^d$ as well as  $\ETbar^u$ and $\ETbar^d$.
However the  convolution integrals have real and imaginary parts. So it is impossible to  solve these equations unambiguously with only two equations in hands.
As a guidance  we can estimate the form factors if we suppose that the
relative phase $\Delta\phi$ between $\HT^u$ and $\HT^d$ equals 0 or 180 degrees. 
Ignoring an overall phase, the form factors are then real and we arbitrarily choose the solution with 
  $\HT^u$ and  $\ETbar^u$ positive.
Fig.~\ref{fig:ud_3}  presents 
$\HT^u$, $\HT^d$, $\ETbar^u$ and $\ETbar^d$
for one kinematic point $(Q^2=2.2~GeV^2,x_B=0.27)$ calculated in this assumption. 
Note the different signs of $\HT^u$ and  $\HT^d$ convolutions and the same sign
of $\ETbar^u$ and $\ETbar^d$.


\section{Conclusion}
Differential cross sections of exclusive $\pi^0$ and $\eta$ electroproduction have been obtained in the few-GeV region at more than 1800 kinematic points in  bins of $Q^2, x_B$, $t$ and $\phi_\pi$. 
Virtual photon structure functions  
$\sigma_U$, $\sigma_{TT}$ and $d\sigma_{LT}$ have been obtained. It is found that $\sigma_U$ and $\sigma_{TT}$ are comparable in magnitude with each other, while $\sigma_{LT}$ is very much smaller than either. 
Generalized form factors of the transversity GPDs $\HT^{\pi,\eta}$ and $\ETbar^{\pi,\eta}$ were directly extracted from the experimental observables
for the first time. It was found that the GPD $\bar E_T$ dominates in  pseudoscalar meson production.
The combined $\pi^0$ and $\eta$ data opens the way for the flavor decomposition of the transversity GPDs. 
Within some simplifying assumptions, the  decomposition has been  demonstrated for the first time.


\section*{Acknowledgments}
The author  thanks G. Goldstein, S. Goloskokov, P. Kroll, J. M. Laget, S. Liuti and A.~Radyushkin  for many informative discussions and making available the results of their calculations. 
This work was supported in part by 
the U.S. Department of Energy and National Science Foundation, 
The Jefferson Science Associates (JSA) operates the Thomas Jefferson National Accelerator Facility for 
the United States Department of Energy under contract DE-AC05-06OR23177. 


\end{document}